\documentclass[epsfig,11pt]{llncs}
\usepackage{amssymb,amsmath,graphicx,color}
\usepackage{amssymb,amsbsy,amsmath,graphicx,color,mathrsfs}
\usepackage{amsfonts}
\usepackage{epsfig,latexsym}
\usepackage{amsfonts}
\usepackage{epsfig,latexsym,graphicx}

\begin{document}
\title{Tests for exponentiality against NBUE alternatives: a Monte Carlo comparison}
\author{M.Z. Anis, Kinjal Basu}
\institute{Indian Statistical Institute\\
203, B.T. Road, Kolkata - 700 108, India\\
Email : zafar@isical.ac.in, bst0707@isical.ac.in}
\maketitle

\begin{abstract}
Testing of various classes of life distributions has been addressed in the literature for more than forty-five years. In this paper, we consider the problem of testing exponentiality (which essentially implies no aging) against positive aging which is captured by the fairly large class of New Better than Used in Expectation (NBUE) distributions. These tests of exponentiality against NBUE alternatives are discussed and compared. The empirical size of the tests is obtained by simulation. Power comparisons for different popular alternatives are done using Monte Carlo simulation. These comparisons are made both for small and large sample sizes. The paper concludes with a discussion in which suggestions are made regarding the choices of the test when a particular alternative is suspected.
\end{abstract}

\section{Introduction}
The exponential distribution is characterized by the lack of memory property; a constant hazard rate and a constant mean residual life function. This is the most commonly used distribution in reliability and survival analysis, primarily because of its mathematical simplicity. The assumption of exponential life times essentially means that a used item is stochastically as good as a new one; hence there is no need to replace a unit which is working. However, this is not always a realistic assumption; and age does have an effect on the residual life time. Positive (negative) aging means that age has an adverse (beneficial) effect, in some probabilistic sense, on the residual life. Hence, it is of interest to check possible departure from exponentiality in the data.

A hierarchy of aging classes has been proposed in the literature to model the effect of aging. One such aging class is the so-called New Better than used in Expectation (NBUE) class, which is defined below.
\\
\\
\textbf{Definition 1}\textit{ : A non-negative random variable X with finite mean $\mu$ is said to possess the NBUE property if $e_F(t) = E_F(X-t|X>t) \leqslant E_F(X)$ i.e}
\[\int_{t}^{\infty}\overline{F}(x)dx \leqslant \mu\overline{F}(t).\]

Important results for this class are discussed in Barlow and Proschan (1975). Specifically the NBUE family of life distributions is important in the study of replacement policies. In fact, Marshall and Proschan (1972) have proved that the average waiting time between any two consecutive failures when no planned replacement policies is adopted is smaller than or equal to the similar quantity when an age replacement policy is adopted if and only if the life distribution is NBUE. 

Furthermore, this average waiting time is the same under these two policies only when the system life is exponentially distributed. Hence, if the average waiting time between the consecutive failure is an important criterion in deciding whether to adopt an age replacement policy over the failure replacement policy for a given system, then a reasonable way to decide would be to test whether the life distribution of the given system is exponential. Rejection of the exponential hypothesis on the basis of the given data would imply as favoring the adoption of age replacement policy.

Doksum and Yandel (1984) and Spurrier (1984) were the first one to review the tests of exponentiality. This was followed up with the works of Ascher (1990) and Henze and Meintanis (2005). Anis and Dutta (2010) discuss tests of exponentiality against IFR alternatives; while Anis (2010) reviews the procedures of testing exponentiality against DMRL alternatives. We contribute to the literature by examining the tests of exponentiality against NBUE alternatives.

This paper is organized as follows. We introduce the different test procedures in Section 2. The results of a Monte Carlo study on the size of the tests is reported in Section 3. We present in Section 4 the power of these tests against the commonly used alternatives to the exponential model. Finally concluding remarks are made in Section 5.

\section{The Test Statistics}
The problem of testing exponentiality against NBUE alternatives is more than three decades old.
Different types of approaches are used in deriving the test statistics. The first test procedure was proposed by Hollander and Proschan (1975) based on a distance measure. de Souza Borges et al. (1984) proposed a test based on the coefficient of variation. Belzunce et al. (2000) had considered the same problem by means of the dispersion of residual lives; while Belzunce et al. (2001) proposed a test based on the right spread order. Fernandez-Ponce et al. (1996) suggested a test based on the right-spread function to characterize different partial orderings between life time distributions. Anis and Mitra (2011) have generalized the Hollander-Proschan statistic to obtain a family of test statistics.

Formally we are concerned with testing the following hypothesis: 
\[\textnormal{H}_0 : F \in \mathscr{E}\;\; \textnormal{v.s H}_1 : F \in \textnormal{NBUE -} \mathscr{E}
\]
based on a random sample $X_1, X_2, \ldots, X_n$ of size $n$ from an absolutely continuous distribution function $F(x) = 1 - e^{-\lambda x}, x \geqslant 0$ and $\lambda > 0$, typically unknown. Let $\overline{F} = 1 - F$ be the corresponding survival function. Let $0 = X_{(0)} \leqslant X_{(1)} \leqslant X_{(2)} \leqslant \ldots \leqslant X_{(n)}$ be the corresponding order statistics. Let $\overline{X}$ denote the sample mean.

We shall now briefly discuss these tests.

\subsection{The First Test Procedure}

Hollander and Proschan (1975) suggested a test statistic by considering the integral
\begin{equation}
\gamma(F) = \int_0^{\infty}\overline{F}(t)\left(e_F(0) - e_F(t)\right)dF(t)
\end{equation}
as a measure of deviation from $H_0$ towards the alternative $H_1$. The sample counterpart to  $\gamma(F)$ is obtained by substituting the empirical distribution function $F_n$ for $F$ in (1) above. On simplification, they obtained

\begin{equation}
K = \frac{1}{n^2}\sum_{i=1}^{n}X_{(i)}\left\{\frac{3n}{2} - 2i + \frac{1}{2}\right\}
\end{equation}
where $X_{(j)}$ denotes the $j$-th order statistic. In order to make the test statistic scale invariant, they suggest using $T_1 = K/\overline{X}$. They have proved that their test is consistent and asymptotically normal. The null hypothesis is rejected for large value of $T_1$.

\subsection{Test due to Koul (1978)}
By definition it follows that $F$ is NBUE if 
\[\int_{0}^{y}\overline{F}(x)dx \geqslant \mu F(y) \qquad 0<y<\infty
\]
where $\mu = \int_0^\infty \overline{F}(y)dy < \infty$. Then define
\begin{equation}
D := \sup_{0 < y < \infty}\left\{\frac{1}{\mu}\int_0^y\overline{F}(x)dx - F(y)\right\}.
\end{equation}

Observe that if $F$ is exponential, then $D = 0$. The farther $D$ is from 0, the more evidence there is for $F \in H_1$. Substituting for $F$ by the empirical distribution function $F_n$ and simplifying Koul (1978) obtained the statistic 
\begin{equation}
T_2 = D_n \equiv D(F_n) = \max_{1\leqslant i \leqslant n}\left\{W_{ni} - \frac{i}{n} \right\}
\end{equation}
where 
\[
W_{ni} = \frac{\sum_{j=1}^{i}(n - j + 1)\left\{X_{(j)} - X_{(j-1)}\right\}}{\sum_{j=1}^{n}(n - j + 1)\left\{X_{(j)} - X_{(j-1)}\right\}}.
\]
The test rejects $H_0$ in favour of $H_1$ if $T_2$ is large and is consistent. The null distribution of this test procedure is derived in Barlow and Doksum (1972).
\subsection{Test based on the Coefficient of Variation}
de Souza Borges et al. (1984) introduced a simple test based on the sample coefficient of variation. They considered the parameter
\begin{equation}
\Delta(F) = \int_0^\infty\overline{F}(t)\left\{\mu - e_F(t)\right\}dt
\end{equation}
as a measure of the deviation from the null hypothesis of exponentiality. They proved the following characterization of exponentiality for life distributions which are continuous NBUE.
\\\\
\textbf{Theorem :}
\textit{Let $F$ be a continuous life distribution and NBUE. A necessary and sufficient condition for F to be exponential is that $CV(F) = 1$, where $CV(F)$ denotes the coefficient of variation of $F$}
\\\\
Using the above characterization theorem, they suggest a test based on the sample coefficient of variation ${S}/{\overline{X}}$, where
\[S^2 = \frac{1}{n}\sum_{i=1}^{n}\left(X_i - \overline{X}\right)^2.
\]
Under $H_0$, ${S}/{\overline{X}}$ is asymptotically normal with mean 1 and variance $1/n$. A large sample level $\alpha$ test rejects $H_0$ in favour of the NBUE property if $T_3 \equiv \sqrt{n}({S}/{\overline{X}} - 1) \leq -z_\alpha$, the standard normal $\alpha$-quantile. (The test rejects $H_0$ in favour of the NWUE property if $\sqrt{n}({S}/{\overline{X}} - 1) > z_\alpha$). They also showed that the test is consistent.

\subsection{Test due to Aly (1990)}
Aly considered the parameter
\[\begin{split}
\gamma_{F} &= \int_0^1(1-y)\left\{1 + \textnormal{log}(1-y)\right\}dF^{-1}(y) \qquad 0 \leqslant y \leqslant 1\\
&= \int_0^\infty\left\{e_F(0) - e_F(t)\right\}dF_(t)
\end{split}
\]
as a measure of NBUE-ness. Substituting for the empirical distribution function and simplifying he obtained
\begin{equation}
T_4 \equiv \gamma(F_n) = \sum_{i=1}^{n}\left\{1 + \textnormal{log}\left(\frac{n-i+1}{n}\right)\right\} \left(\frac{n-i+1}{n}\right)\left\{\frac{X_{(i)} - X_{(i-1)}}{\overline{X}_n}\right\}
\end{equation}

He has proved that $\sqrt{n}T_4$ is asymptotically standard normal; and is consistent. The test rejects the nul hypothesis of exponentiality at (approximate) level $\alpha$ if
\[\frac{\sqrt{n}\left\{\gamma(F_n) - \lambda_n\overline{X}_n\right\}}{\sigma_n\overline{X}_n} \geqslant z_{1-\alpha}
\]
where
\[
\begin{split}
\lambda_n &= 1 + \frac{1}{n}\sum_{j=1}^{n}\textnormal{log}\left(1 - \frac{j-1}{n}\right),\\
\sigma_n^2 &= \frac{1}{n}\sum_{j=1}^{n}\left\{1 + \textnormal{log}\left(1 - \frac{j-1}{n}\right)\right\}^2,\\
\end{split}
\]
and $z_{1 - \alpha}$ is the $(1-\alpha)$ quantile of the standard normal distribution.

\subsection{Test based on a Quantile Dispersion Measure}
Fernandez-Ponce et al. (1996) proposed the test statistic $T_5$ based on the Right Spread (RS) function. The RS function of the random variable $X$ is defined as 
\begin{equation}
S_X^+(u) = E\left((X - F_X^{-1}(u))^+\right) = \int_{F_X^{-1}(u)}^{\infty}\overline{F}_X(x)dx
\end{equation}
where $x^+ = \max \{x,0\}$ and $F^{-1}(x)$ is the left continuous inverse (quantile function), 
$Q_{X}(u) \equiv F^{-1}(x) = \inf\{x : F_X(x) \geqslant u\};\;\; \forall u \in (0,1)$.
\\\\
\textbf{Definition 2}\textit{ : F is more new better than used in expectation than G, (denoted by F $\stackrel{NBUE}{<}$ G) if}
\[\frac{e_F\{F^{-1}(x)\}}{e_G\{G^{-1}(x)\}} \leqslant \frac{\mu_F}{\mu_G}, \qquad x \in [0,1].
\]
It can be shown than $F$ is $NBUE$ if and only if F $\stackrel{NBUE}{<}$ G , where $\overline{G} = e^{-x}$.
\\\\
Their test is motivated by considering the following integral as a measure of deviation, for a given $F_x$, from $H_0$ to $H_1$ :
\begin{equation}
\Psi(F) = \int_0^1\left\{1 - L_{S^+}^u(0)\right\}du
\end{equation}
where $L_{S^+}^u(0)$ is the length of the segment determined by the secant to $S_X^+(u)$ from $U=0$ to $U=u$ in the x-axis. Making the change $u = F_X(t)$ one has
\begin{equation}
\Psi(F) = \int\left\{1 - L_{S^+}^{F(t)}(0)\right\}dF(t).
\end{equation}
The parameter $\Psi(F)$ can be viewed as follows. Define $D(t) = 1 - L_{S^+}^{F(t)}(0)$. Now, $D(t) = 0 \;\;\forall t \in \mathbb{R} \Leftrightarrow H_0$ is true. To show this equivalence observe that,
\begin{equation}
\Psi(F) = E_F\left(1 - L_{S^+}^{F(t)}(0)\right).
\end{equation}
So $\Psi(F)$ is simply an average value of the deviation from $H_0$ towards $H_1$. Note that $\Psi(F) \geqslant 0\;\;\forall\;\; F \in H_0 \cup H_1$ and $\Psi(F) = 0$ if and only if $F \in H_0$.

The sample analogue of the parameter $\Psi(F)$ is the basis for their proposed test statistic. Given a random sample of size $n$, they obtain the test statistic $T_5$, by replacing $F(t)$ by its natural estimates, namely the empirical distribution function. Thus Fernandez-Ponce et al. (1996) obtain
\begin{equation}
T_5 \equiv \widehat{\Psi}(F_n) = 1 - \frac{1}{n}\sum_{i=1}^{n-1}\left\{\frac{i}{n}\frac{\sum_{k=1}^{n}(n-k+1)\left(X_{(k)} - X_{(k-1)}\right)}{\sum_{k=1}^{i}(n-k+1)\left(X_{(k)} - X_{(k-1)}\right)}\right\}
\end{equation}
with the assumption that $X_{(0)} = 0$.

Fernandez-Ponce et al. (1996) state that $T_5$ is asymptotically unbiased and converges almost surely to $\Psi(F)$. Though the exact distribution of $T_5$ could not be found, the critical points for sample size $n = 3\;(1)\;20\;(5)\;35$ for different significance levels have been tabulated in Fernandez-Ponce et al. (1996). 

\subsection{Test based on Dispersion of Residual Life}

Belzunce et al. (2000) proposed a new test for $H_0$ based on dispersion of residual life. Before describing their test, we need to give some background definitions.
\\\\
\textbf{Definition 3} :\textit{(Hickey (1986)) A random variable $Y$ is more dispersed in dilation than a random variable X (denoted by $X <_{dil} Y$) if}
\begin{equation}
E\left(\phi(X -E(X))\right) \leqslant E\left(\phi(Y -E(Y))\right)
\end{equation}
\textit{for every convex function $\phi$.}
\\\\
Belzunce et al. (1997) have proved that 

\begin{equation}
X\;\textnormal{is NBUE if and only if}\;\; X_t >_{dil} Y, \;\; \forall t > 0
\end{equation}
 where $X_t \equiv \{X - t|X>t\}$ is the remaining life at age $t$ and $Y$ is an exponential random variable with mean equal to $E(X)$, i.e. $E(Y) = E(X)$.

Using the characterization in (13) above, Belzunce et al. (2000) consider
\begin{equation}
\bigtriangleup_{NBUE}(X) = \int_{\mathbb{R}}\overline{F}^{2}_X(t)\left\{\bigtriangleup_{dil}(X_t,Y)\right\}dF_X(s)
\end{equation}
as a measure of departure from $H_0$ to $H_1$, where
\begin{equation}
\bigtriangleup_{dil}(X,Y) = \int_{l_X}^{\infty}\overline{F}_X^2(t)dt - E(X) - \int_{l_Y}^{\infty}\overline{F}_Y^2(t)dt + E(Y)
\end{equation}
with $l_X$ and $l_Y$ being the smallest values of the supports of $F_x$ and $F_y$, which are assumed to be finite; and where $Y$ is an exponential distribution with mean equal to $E(X)$.

Now, $\bigtriangleup_{NBUE}(X) = 0$ under the null hypothesis of exponentiality, and under $H_1$, $\bigtriangleup_{NBUE}(X) > 0$. Based on a random sample of size $n$, the empirical analogue is given by
\begin{equation}
\bigtriangleup_{NBUE}(n) \equiv \frac{1}{n^4}\sum_{i=0}^{n-2}n(n-i)^2\left(\lambda_n(i) + \frac{\sum_{i=1}^{n}X_{(i)}}{2n}\right)
\end{equation}
where
\begin{eqnarray}
\lambda_n(i) &\equiv& \frac{1}{(n-i)^2}\bigtriangledown_i^n,\\
\bigtriangledown_i^n &\equiv& \sum_{\alpha = i+1}^{n}\delta_{i_\alpha}X_{(\alpha)},\\
\delta_{i_\alpha} &=& n - 2\alpha + i + 1.
\end{eqnarray}

The null hypothesis of exponentiality is rejected for large values of $\bigtriangleup_{NBUE}(n)$. Since $\bigtriangleup_{NBUE}(n)$ is not scale invariant, Belzunce et al. (2000) suggest using

\begin{equation}
T_6 \equiv \bigtriangleup_{NBUE}^*(n) = \frac{\bigtriangleup_{NBUE}(n)}{\overline{X}_n}.
\end{equation}

They have derived the exact distribution of $T_6$ and have shown that $T_6$ is asymptotically normal. Under $H_0$, the limiting distribution of $\sqrt{45n}T_6$ is standard normal. They have proved that the test is consistent and the critical points based on the exact distribution is obtained for $n = 2\;(1)\;20\;(5)\;60$. For large values of $n$, the null hypothesis of exponentiality is rejected if $\sqrt{45n}T_6 > Z_\alpha$.

\subsection{Test based on the Right Spread Order}

Belzunce et al. (2001) suggested a new test procedure based on the right spread order.

Recall that the Right Spread Function has already been defined in sub-section 2.5 above. It should be noted that the right spread function can be considered as a measure of spread to the right of every quantile $Q_X(u)$.

A random variable $X$ is said to be less than $Y$ in the right spread order (denoted by $X \leqslant_{RS} Y$) if
\begin{equation}
 E\left\{(X - F_X^{-1}(u))^+\right\} \leqslant   E\left\{(Y - F_Y^{-1}(u))^+\right\} \qquad \forall u \in (0,1).
\end{equation}

It is interesting to note that the right spread function of a random variable $X$ is also related to its mean residual life function $e_F(t) = E(X-t|X>t)$ by the relationship
\begin{equation}
 E\left\{(X - F_X^{-1}(u))^+\right\} = (1-u)e_F(F_X^{-1}(u)).
\end{equation}
Then it is easy to show that 
\begin{equation}
X \in NBUE \Leftrightarrow X \leqslant_{RS} Y
\end{equation}
where $Y$ is an exponentially distributed random variable such that $E(Y) = E(X)$.
\\\\
Belzunce et al. (2001) use the above characterization in (23) above to construct a family of tests for NBUE aging class. In fact their proposed test statistic is similar to their test statistic $\bigtriangleup_{RS}^\alpha(X,Y)$ proposed to test the right spread order; i.e.
\[H_0' : X =_{RS}Y\;\; \textnormal{v.s.}\;\; H_1' : X \leqslant_{RS} Y.
\]
More specifically for testing the null hypothesis of exponentiality against NBUE alternatives, they suggest taking $\bigtriangleup_{RS}^\alpha(X,Y)$ where $Y$ is an exponential random variable with $E(Y) = E(X)$, as a measure of departure from $H_0$ towards $H_1$. Then it can be shown that

\begin{equation}
\bigtriangleup_{NBUE}^\alpha \equiv \bigtriangleup_{RS}^\alpha(X,Y) = \frac{1}{6}E(X)(1 - \alpha)(2-\alpha) - \int_0^1J_\alpha(p)\int_{F_X^{-1}(p)}^{\infty}\overline{F}_X(x)dx\;dp
\end{equation}
where
\[J_\alpha(p) =
\begin{cases}
p\left(\frac{1}{\alpha} - 1\right) & 0\leqslant p \leqslant \alpha\\
1 - p & \alpha \leqslant p \leqslant 1.
\end{cases}
\]
Then under the null hypothesis of exponentiality $\bigtriangleup_{NBUE}^\alpha(X) = 0$,  but under $H_1$, $\bigtriangleup_{NBUE}^\alpha(X) > 0$.  Based on a random sample, the sample analogue of the measure in (25) above is given by
\begin{equation}
\widehat\bigtriangleup_{NBUE}^\alpha(X_1,.., X_n) = \frac{1}{6}\overline{X}(1-\alpha)(2 - \alpha) - \frac{1}{n}\sum_{i=1}^{n}\left\{L_\alpha\left(\frac{i}{n}\right) - J_\alpha\left(\frac{i}{n}\right)\left(1 - \frac{i-1}{n}\right)\right\}X_{(i)} 
\end{equation}
and 
\[L_\alpha\left(\frac{i}{n}\right) =
\begin{cases}
\frac{1}{2}\left(\frac{1}{\alpha} - 1\right)\left\{\frac{i^2}{n^2} + \frac{i}{n^2}\right\} & \textnormal{if} \;\;\frac{i}{n} \leqslant \alpha\\
-\frac{i^2}{2n^2} + \frac{i}{n}\left(1 - \frac{1}{2n}\right) + \frac{l^2}{n^2}\frac{1}{2\alpha} + \frac{l}{n}\left(\frac{1}{2\alpha n} - 1\right) & \textnormal{if}\;\; \frac{i}{n} > \alpha\\
\end{cases}
\]
where $\frac{l}{n} \leqslant \alpha, \frac{l+1}{n} > \alpha$.
\\\\

Since this statistic is not scale invariant, they suggest using 
\begin{equation}
T_7 = \frac{\widehat\bigtriangleup_{NBUE}^\alpha(X_1,.., X_n)}{\overline{X}}.
\end{equation}

Belzunce et al. (2001) have found the exact distribution of $T_7$, under the null hypothesis of exponentiality. Further, they have proved that $T_7$ is  asymptotically normal and under $H_0$, the limiting distribution of 
\[\frac{1}{\alpha - 1}\sqrt{\frac{45n}{1 + 2\alpha - 2\alpha^2}} T_7 \;\;\textnormal{is}\;\; N(0,1).
\]

Thus the null hypothesis of exponentiality is rejected at level $\alpha$ if for large $n$, $\frac{1}{\alpha - 1}\sqrt{\frac{45n}{1 + 2\alpha - 2\alpha^2}} T_7 > z_\alpha$.  They have also proved that the test is consistent.

\subsection{Test based on Moment Inequality derived from comparing Life with its Equilibrium Form}

It should be noted that associated with the life distribution $X$ is the notion of ``random remaining life" at age $t$, denoted by $X_t$. It is easy to show that $X_t$ has a survival function $\overline{F}_t(x) = \overline{F}(x+t)/\overline{F}(t), x,t \geqslant 0$. It is well known that, see for example Ross (2003), that $X_t$ converges weakly to a non-negative random variable $\widetilde{X}$ with survival function $\overline{W}_F(x) = \frac{1}{\mu}\int_x^\infty \overline{F}(u)du, x\geqslant 0$.
\\

In an interesting paper Mugdadi and Ahmad (2005) showed that the notion that $X$ is NBUE is equivalent to the notion that $\widetilde{X} \stackrel{st}{\leqslant} X$;   where $\widetilde{X}$ denotes the equilibrium life, and  $\stackrel{st}{\leqslant}$ denotes less than or equal to in the usual stochastic order.

Then by Theorem 3 of Mugdadi and Ahmad (2005), $H_0$ can be tested by comparing $X$ and the equilibrium life $\widetilde{X}$. Since $\widetilde{X} \stackrel{st}{\leqslant} X$ means that,
\begin{eqnarray}
&&\int_x^\infty\overline{F}(u)du \leqslant u\overline{F}(x)\\
&\Rightarrow& \int_0^\infty\left(\int_x^\infty\overline{F}(u)du\right)dF(x) \leqslant \frac{\mu}{2}\\
&\Rightarrow&\int_0^\infty\overline{F}^2(x)dx \leqslant \frac{\mu}{2}
\end{eqnarray}
Hence as a measure of departure from $H_0$, we may consider 
\begin{equation}
T_8 = \frac{1}{\mu}\left\{\frac{E(X_1)}{2} - E\left(\min(X_1,X_2)\right)\right\}.
\end{equation}
Then $T_8$ can be estimated by
\begin{equation}
\widehat{T}_8 = \frac{1}{\overline{X}}\left(\frac{1}{n(n-1)}\sum_{i=1}^n\sum_{j\neq i}\left\{\frac{X_1}{2} - \min(X_i,X_j)\right\}\right).
\end{equation}
Using U-statistics theory, Mugdadi and Ahmad (2005) show that $\sqrt{n}(\widehat{T}_9 - T_8)$ is asymptotically normal with mean 0 and variance 
\begin{equation}
\sigma^2 = \textnormal{Var}\left(\frac{X_1}{2} + \frac{\mu}{2} - 2\int_0^XxdF(x) - 2X_1F(X_1)\right).
\end{equation}
Then under $H_0, \sigma^2 = 1/12$. The test rejects $H_0$ for large value of the statistic $\widehat{T}_8$.

\subsection{A Generalization of the Hollander-Proschan Type Test}

Anis and Mitra (2011) proposed a family of test statistics by generalizing the Hollander and Proschan (1975) approach. Essentially they consider the measure
\begin{equation}
\gamma_j(F) = \int_0^\infty\left(\overline{F}(t)\right)^j\left\{e_F(0) - e_F(t)\right\}dF(t)
\end{equation}
as a measure of departure from $H_0$ towards the alternative $H_1$. The sample analogue of $\gamma_j(F)$ reduces to 
\begin{eqnarray}
\gamma_j(F) &=& \frac{1}{j}\sum_{k=1}^{n}X_{(k)}\left\{\left(\frac{n-k+1}{n}\right)^{j+1} - \left(\frac{n-k}{n}\right)^{j+1} - \frac{1}{n(j+1)}\right\}
\end{eqnarray}
Clearly $\gamma_j(F_n)$ is an L-statistic, being a linear combination of order statistics. As $\gamma_j(F_n)$ is not scale invariant, Anis and Mitra (2011) suggest using $T_0 = \gamma_j^*(F_n) \equiv \gamma_j(F_n)/\mu$, where $\mu = E(X)$. The null hypothesis is rejected for large values of $\gamma_j^*(F_n)$. They proved that $\gamma_j^*(F_n)$ is asymptotically normal and the test is consistent. Based on the sample observation, the scale invariant statistic $\gamma_j^*(F_n)$ is estimated by $\widehat{T}_0 = \widehat{\gamma_j^*}(F_n) = \gamma_j^*(F_n)/\overline{X_n}$, where $\overline{X}_n = \sum_{i=1}^{n}X_i/n$.

Very recently, Anis and Basu (2011) have obtained the exact sampling distribution of $\gamma_j^*(F_n)$. They have also given a set of critical values for the commonly used significance levels and sample sizes $n = 2\;(1)\;25\;(5)\;100$. They conclude from their table that the convergence to normality is extremely slow.

\section{Size of Different Tests}
In a large simulation study, we calculate the empirical size of the tests under consideration; i.e, the rejection probability under $H_0$. We do this for different sample sizes and for nominal size (level of significance) of 5\%. To calculate the empirical size of the tests we simulate $n$ observations from an exponential distribution and compute the respective test statistic and check whether this particular realization of the test statistic accepts or rejects the null hypothesis of exponentiality. This procedure is repeated for $10^5$ times and the proportion of times that the test statistic rejects the null hypothesis is observed. Thus we estimate the size for the different tests. The standard errors of the estimated sizes are less than or equal to $\sqrt{0.25/10^5} = 0.0016$. All simulations were done using MATLAB 7.8 on PC platform.

\subsection{Size for small sample}

Hollander and Proschan (1975) refers to Barlow (1972) for the small sample critical values. Fernandez-Ponce et al. (1996) and more recently Belzunce et al. (2000) provide the critical values for small sample sizes. Anis and Basu (2011) have provided the critical values for the Anis and Mitra (2011) test $T_0$. Anis and Basu (2011) have also provided the critical values for the Hollander and Proschan (1975) test statistic based on the exact distribution which they have obtained. Table 1 shows the simulated sizes for these four tests for $n = 5\;(1)\;15$  and for a nominal significance level $\alpha = 5\%$. Also included are the simulated sizes for the Anis-Mitra test $T_0$ for $j=1$. It is seen that the Anis-Mitra test with $j = 0.25$ appears to be very conservative being always less than 0.05. The test $T_5$ due to Fernandez-Ponce et al. (1996) overshoots the nominal size.

\subsection{Size for moderate sample size}
We again compare the above mentioned four tests procedures for moderate sample sizes. We consider $n = 16\;(1)\;20\;(5)\;30$. Similar trends are seen. The Anis-Mitra (2011) test $T_0$ is most conservative; while $T_5$ due to Fernandez-Ponce et al. (1996) overshoots the nominal size most of the time, as seen in Table 2.

\subsection{Size for large sample size}

For large sample sizes, most of the test statistics are normally distributed. However, whenever the exact critical values are available, we have used the exact value. The test proposed by Fernandez-Ponce et al. (1996) has a very complicated statistic and its asymptotic distribution has not been derived. Hence this test procedure has not been considered for large sample.

From Table 3, we see that the test $T_2$ due to Koul (1978) and the test $T_3$ based on the coefficient of variation due to de Souza Borges et al. (1984) are very conservative. The test $T_7$ based on the right spread order given by Belzunce et al. (2001) is also conservative. The test $T_6$ based on dispersion of residual lives constructed by Belzunce et al. (2000) overshoots the nominal size for most of the sample sizes considered. The test $T_4$ due to Aly (1990) performs poorly and overshoots the nominal size for all sample sizes. The test $T_8$ based on the equilibrium distribution given by Mugdadi and Ahmad (2005) also exceeds the nominal size. Finally the tests $T_0$ given by Anis-Mitra (2011) and Hollander Proschan (1975) $T_1$ perform in a mixed manner; for some sample sizes the tests are conservative while sometime they overshoot the nominal size.

\section{Power of the Tests}

The power of the tests under comparison has been evaluated by Monte Carlo simulation. The power estimates were calculated as the proportion of $10^5$ Monte Carlo samples that resulted in rejection of the null hypothesis of exponentiality at significance level $\alpha$ of $5\%$ for the alternative distributions considered. The standard errors of the estimated probabilities are bounded above by $\sqrt{0.25/10^5} = 0.0016$. As mentioned in the previous section, the simulations were done using MATLAB 7.8 on PC platform. The power of the tests is simulated from the following distributions:
\begin{itemize}
\item The Weibull distribution with density $\theta x^{\theta -1}exp(x^\theta), \theta \geqslant 1, x \geqslant 0$, denoted by $W(\theta)$;
\item The Gamma distribution with density $\{\Gamma(\theta)\}^{-1}x^{\theta - 1}exp(-x), \theta \geqslant 1, x \geqslant 0$, denoted by $\Gamma(\theta)$;
\item The linear failure rate distribution with density $(1 + \theta x)exp(-x - \theta x^2/2), \theta \geqslant 0, x \geqslant 0$, denoted by $LFR(\theta)$.
\end{itemize} 
Note that these distributions are generally used as alternatives to the exponential model. It is easy to show that the Weibull and gamma distributions belong to the NBUE class for $\theta > 1$; while the LFR distribution is NBUE for $\theta > 0$. The null distribtuion is obtained at $\theta_0 = 1$ for the first two; whereas the LFR distribution reduces to the exponential distribution at $\theta_0 = 0$.

In a large simulation study, we simulated the power for different alternatives for sample sizes $n = 5\;(5)\;100$. However, for brevity we shall report our findings for $n=5\;(5)\;30\;(10)\;100$. We simulated the powers for Weibull alternatives for shape parameter $\theta = 1.10\;(0.10)\;1.50$. For the gamma distribution, the shape parameter $\theta$ selected were $\theta = 1.20\;(0.20)\;2.00$. Finally, for the LFR distribution, the shape parameter $\theta$ chosen were $\theta = 0.25\;(0.25)\;1.25$. It should be noted that for these parameter values the distributions are NBUE. Hence a ``good" test should enjoy high power.

It may be pointed out that Anis and Mitra (2011) have observed an error in the computational formula for the Hollander-Proschan test statistic. Anis and Basu (2011) have made a comparative study of the Hollander-Proschan statistic $T_1$ and the Anis-Mitra statistic $T_0$ with $j=1$. It is further reinforced from the simulation results in Table 4-6, that we should use the correct statistic as given by Anis and Mitra (2011). Though the difference is of the order of $1/n^2$, it is appreciable for small sample sizes, as expected.

\subsection{Power for small sample size}

We are able to compare only four test procedure when the sample size is small. These tests are due to Hollander and Proschan(1975) $T_1$; Fernandez-Ponce et al. (1996) $T_5$; Belzunce et al. (2000) $T_6$ and Anis and Mitra (2011) $T_0$. For small sample comparative study we chose the sample size $n=5\;(5)\;25$. Tables 4-6 show the simulated power of these tests against the Weibull, Gamma and the LFR alternatives. We observe from these tables that the performance of each of these tests is poor when the NBUE property is marginal. This is expected because the test can easily confuse and consider the observations to have come from an exponential distribution. 

From Tables 4 and 5 it is observed that if the Weibull or Gamma alternatives are suspected then it best to use the Anis-Mitra test $T_0$ for very small ($n \leqslant 10$) sample size; while it is preferable to use the test $T_5$ due to Fernandez-Ponce et al. (1996) for moderate sample size ($15 \leqslant n \leqslant 25$). It is also observed that it is better to use the Anis-Mitra test with $j=1$ instead of the original Hollander-Proschan test $T_1$. The test $T_6$ due to Belzunce et al. (2000) comes a poor third. 

For the LFR alternatives, it is observed from Table 6, that the Anis-Mitra test $T_0$ with $j = 0.25$ performs best for all sample sizes $n \leqslant 25$. The performance of test $T_5$ due to Fernandez-Ponce et al. (1996) and $T_6$ due to Belzunce et al. (2000) based on dispersion of residual lives is comparable.

\subsection{Power for large sample size}

We have compared the performance of eight of the tests against Weibull, Gamma and LFR alternatives for different values of the shape parameter $\theta$ and for sample size $n = 30\;(5)\;100$. For brevity of space we give in Table 7-9 the power comparisons for sample sizes $n = 30, 40, 50, 75$ and 100 for different alternatives. The test $T_5$ due to Fernandez-Ponce et al. (1996) has not been considered as neither the distribution of the test statistic is derived nor the critical values (based on simulation) are provided.

It is seen from Table 7 and 8 that for Weibull and gamma alternatives, the Anis-Mitra test $T_0$ with $j = 1$ performs best. The performance of the original Hollander-Proschan test $T_1$ is comparable, though the power figures are smaller (in view of the inherent error in the test statistic). The performance of the Anis-Mitra test $T_0$ with $j = 0.25$ is also very good. For Weibull alternative, the test due to Mugdadi and Ahmad (2005) $T_8$ is worst. For gamma alternatives, the worst performance is exhibited by the tests due to Aly (1990) $T_4$ and Mugdadi and Ahmad (2005) $T_8$. As an example, when the alternative is Weibull with $\theta = 1.5$, with a moderately large enough sample of size $n = 30$, the Anis-Mitra test with $j=1$ is able to take the correct decision 81.87\% of the time. The corresponding figure for the Mugdadi and Ahmad test $T_8$ is mere 30.73\% only. Even with a large sample of size $n = 100$, the Mugdadi and Ahmad test $T_8$ is able to take the correct decision only $77.28\%$ of the time when the scale parameter is $\theta = 1.3$, the corresponding figures for the Hollander-Proschan test $T_1$ is 93.17\%, while for the Anis-Mitra test $T_0$ with $j = 1$ is 95.44\% and for $j = 0.25$, it is 92.84\%.

Similarly when the Gamma parameter $\theta = 2$, (a moderate NBUE property) the test $T_4$ due to Aly (1990) is able to take correct decision only 1\% of the time with a sample size of $n = 30$. In contrast the test $T_8$ due to Mugdadi and Ahmad (2005) has a success rate of $24.39\%$. For these parameter values, the test $T_0$ with $j = 1$ shows the estimated power as $79.04$. 

For the LFR alternatives, it is seen that the best performance is exhibited by the test $T_4$ due to Aly(1990). It is a clear winner. The worst performance is shown by $T_8$ due to Mugdadi and Ahmad (2005).
\\\\
\textbf{General Remarks :} It is seen that the test $T_{0;1}$ due to Anis and Mitra (2011) with $j = 1$ out performs the test $T_1$ due to Hollander and Proschan for all sample sizes and for all alternatives considered.

\section{Concluding Remarks}

In this work the performance of several tests for exponentiality against NBUE alternatives has been studied for different sample sizes. We shall summarize the major findings.

It is better to use the test $T_0 (j=1)$ due to Anis and Mitra (2011) instead of the asymptotically equivalent test $T_1$ due to Hollander and Proschan (1975).

For small and moderate sample sizes, the test $T_0 (j = 0.25), T_0 (j = 1)$ and $T_1$ are very conservative. For large sample sizes the tests $T_2$ and $T_3$ are very conservative. On the other hand test $T_4, T_6$ and $T_8$ overshoots the minimal size.

For very small sample size ($n \leqslant 10$) it is better to use $T_0 (j = 0.25)$ due to Anis and Mitra (2011) if the alternative is suspected to be Weibull or Gamma. If the sample size is small $(15 \leqslant n \leqslant 25)$, then the performance of the test $T_5$ due to Fernandez-Ponce et al. (1996) is best and is recommended for use. However, if the sample size is small and the suspected alternative is LFR, then the best performance is exhibited by the test $T_0 (j = 0.25)$ due to Anis and Mitra (2011).

If the sample size is large ($n \geqslant 30$) and the alternative is suspected to be Weibull or gamma alternatives, then it is recommended that $T_0 (j = 1)$ due to Anis and Mitra (2011) be used. The worst performance in this case is exhibited by $T_4$ due to Aly (1990) and $T_8$ due to Mugdadi and Ahmad (2005). In contrast if the alternative is suspected to be LFR, then the best test to use is $T_4$ due to Aly (1990). Even in this case the performance of the test $T_8$ due to Mugdadi and Ahmad (2005) is worst. Hence the test proposed by Mugdadi and Ahmad (2005) should be avoided.

\begin{table}[h]
\footnotesize{
\caption{Size for small sample size}
\begin{center}
\begin{tabular}[c]{|c|c|c|c|c|c|c|}
\hline
Sample&\multicolumn {3}{|c|}{Anis and Mitra (2011)} & Hollander and &  Fernandez-Ponce & Belzunce \\
size&\multicolumn {3}{|c|}{$T_0(j)$}&Proschan (1975)&et al. (1996)&et al. (2000)\\
$n$&$j=0.25$&$j=0.50$&$j=1$&$T_1$&$T_5$&$T_6$\\
\hline
5&4.87&4.92&4.99&5.09&4.50&4.69\\
6&4.89&4.86&4.94&4.82&5.48&4.81\\
7&4.97&4.97&5.01&5.22&5.23&5.06\\
8&4.90&4.93&4.90&5.00&5.06&4.71\\
9&4.87&4.89&4.85&4.83&5.03&5.10\\
10&4.93&4.84&4.83&4.88&4.86&4.46\\
11&4.46&4.45&4.62&4.30&4.65&5.13\\
12&4.83&4.83&4.91&4.38&6.11&5.49\\
13&4.85&4.75&4.90&4.44&5.67&4.76\\
14&4.94&5.03&5.02&4.64&4.41&4.82\\
15&4.87&4.83&4.98&4.62&5.76&4.73\\
\hline
\end{tabular}
\end{center}
}
\end{table}

\begin{table}[h]
\footnotesize{
\caption{Size for moderate sample size}
\begin{center}
\begin{tabular}[c]{|c|c|c|c|c|c|c|c|c|c|}
\hline
Sample&\multicolumn {3}{|c|}{Anis and Mitra (2011)} & Hollander and &  Fernandez-Ponce & Belzunce \\
size&\multicolumn {3}{|c|}{$T_0(j)$}&Proschan (1975)&et al. (1996)&et al. (2000)\\
$n$&$j=0.25$&$j=0.50$&$j=1$&$T_1$&$T_5$&$T_6$\\
\hline
16&4.72&4.56&4.82&4.92&5.88&4.90\\
17&5.27&5.17&5.28&4.22&5.34&5.05\\
18&4.67&4.80&4.77&4.64&4.51&4.73\\
19&4.60&4.73&4.80&4.83&5.31&4.91\\
20&4.96&4.88&4.84&4.63&5.22&5.23\\
25&4.94&5.02&5.06&4.70&5.86&4.90\\
30&4.81&4.82&5.02&4.86&5.62&5.15\\
\hline
\end{tabular}
\end{center}
}
\end{table}

\begin{table}[h]
\scriptsize{
\caption{Size for Large sample size}
\begin{center}
\begin{tabular}[c]{|c|c|c|c|c|c|c|c|c|c|c|}
\hline
&\multicolumn {3}{|c|}{Anis and Mitra} & Hollander \& & Koul & de Souza& Aly& Belzunce & Belzunce&Mudgadi \&  \\
Sample&\multicolumn {3}{|c|}{ (2011)}&Proschan&(1978)&Borges et al.&(1990)&et al.&et al.&Ahmed\\
Size&\multicolumn {3}{|c|}{$T_0(j)$}&(1975)&&(1984)&&(2000)&(2001)&(2005)\\
$n$&$j=0.25$&$j=0.50$&$j=1$&$T_1$&$T_2$&$T_3$&$T_4$&$T_6$&$T_7$&$T_8$\\
\hline
35&5.12&5.16&5.19&4.64&2.48&3.02&6.63&5.19&4.72&5.06\\
40&5.19&5.33&5.07&4.75&2.78&3.05&5.92&4.69&4.85&5.51\\
45&5.25&5.25&5.32&4.86&3.06&3.18&5.99&5.07&4.96&5.20\\
50&4.85&5.03&4.70&4.92&3.03&3.26&6.12&5.14&5.00&5.28\\
55&5.11&5.14&4.97&5.08&3.15&3.48&6.22&5.04&5.07&4.82\\
60&5.26&5.34&5.39&4.90&3.29&3.48&6.36&4.99&4.98&5.08\\
65&4.95&4.86&4.75&4.88&3.19&3.44&5.93&6.63&4.82&5.02\\
70&4.88&4.80&4.62&5.30&3.36&3.76&6.35&7.29&5.23&5.25\\
75&4.77&4.75&4.69&5.31&3.61&3.76&6.28&7.06&5.45&4.91\\
80&4.96&4.87&4.81&5.04&3.80&3.89&6.10&6.76&4.91&5.63\\
85&4.66&4.65&4.80&4.74&3.16&3.56&6.07&6.48&4.73&5.14\\
90&5.20&5.10&5.02&4.68&3.41&3.58&5.79&6.16&4.56&4.82\\
95&5.12&5.29&5.27&4.63&3.13&3.73&5.73&6.28&4.70&5.08\\
100&5.39&5.23&5.21&4.77&3.64&3.66&5.42&6.35&4.99&5.02\\
\hline
\end{tabular}
\end{center}
}
\end{table}

\begin{table}[h]
\footnotesize{
\caption{Estimated power for the Weibull alternative for small sample size}
\begin{center}
\begin{tabular}[c]{|c|c|c|c|c|c|c|c|}
\hline
&Sample&\multicolumn {3}{|c|}{Anis and Mitra (2011)} & Hollander and &  Fernandez-Ponce & Belzunce \\
Weibull($\theta$)&size&\multicolumn {3}{|c|}{$T_0(j)$}&Proschan (1975)&et al. (1996)&et al. (2000)\\
&$n$&$j=0.25$&$j=0.50$&$j=1$&$T_1$&$T_5$&$T_6$\\
\hline
&5&7.00&6.99&7.05&7.11&6.73&7.28\\ 
&10&8.51&8.63&8.70&8.82&5.86&8.36\\ 
$\theta = 1.10$&15&10.01&10.15&10.29&8.88&12.2&9.54\\ 
&20&11.12&11.21&11.35&10.56&12.64&11.22\\ 
&25&11.73&12.02&12.29&11.71&17.26&11.34\\ 
\hline
&5&9.39&9.48&9.42&9.59&8.14&8.43\\ 
&10&12.68&12.74&12.83&13.09&9.71&12.67\\ 
$\theta = 1.20$&15&16.50&16.50&17.07&16.47&20.56&15.59\\ 
&20&21.05&21.45&22.34&19.96&24.18&18.95\\ 
&25&23.06&23.71&24.65&23.74&33.29&21.86\\ 
\hline
&5&11.73&11.95&11.98&12.19&11.38&11.83\\ 
&10&18.53&18.54&18.9&19.14&14.76&17.68\\ 
$\theta = 1.30$&15&26.92&27.66&28.38&25.42&32.43&23.64\\ 
&20&31.90&33.01&34.68&33.15&37.90&29.59\\ 
&25&38.21&39.66&41.36&40.16&51.58&36.20\\ 
\hline
&5&13.84&13.88&14.13&14.32&13.92&14.00\\ 
&10&25.90&26.24&27.28&27.67&21.03&24.25\\ 
$\theta = 1.40$&15&35.88&36.82&38.05&37.58&44.14&33.59\\ 
&20&46.62&48.06&50.29&48.52&53.46&44.16\\ 
&25&55.63&57.97&60.00&58.01&67.87&52.29\\ 
\hline
&5&17.52&17.86&18.04&18.28&16.98&16.72\\ 
&10&33.61&34.44&35.72&36.14&28.87&31.79\\ 
$\theta = 1.50$&15&47.59&49.50&51.06&48.60&57.02&44.48\\ 
&20&59.77&61.60&63.94&62.78&68.89&57.98\\ 
&25&70.49&72.23&74.13&73.17&81.84&66.87\\ 
\hline
\end{tabular}
\end{center}
}
\end{table}

\begin{table}[h]
\footnotesize{
\caption{Estimated power for the Gamma alternative for small sample size}
\begin{center}
\begin{tabular}[c]{|c|c|c|c|c|c|c|c|}
\hline
&Sample&\multicolumn {3}{|c|}{Anis and Mitra (2011)} & Hollander and &  Fernandez-Ponce & Belzunce \\
Gamma($\theta$)&size&\multicolumn {3}{|c|}{$T_0(j)$}&Proschan (1975)&et al. (1996)&et al. (2000)\\
&$n$&$j=0.25$&$j=0.50$&$j=1$&$T_1$&$T_5$&$T_6$\\
\hline
&5&7.41&7.54&7.61&7.68&6.37&6.47\\ 
&10&8.72&8.77&8.86&8.99&5.98&8.04\\ 
$\theta = 1.20$&15&10.66&10.76&10.98&9.85&13.23&9.55\\ 
&20&11.49&11.48&12.15&10.95&14.69&11.16\\ 
&25&12.38&12.77&13.25&12.54&19.81&11.86\\ 
\hline
&5&9.69&9.82&10.08&10.24&8.83&8.93\\ 
&10&13.71&13.76&14.31&14.52&10.13&12.50\\ 
$\theta = 1.40$&15&17.17&17.99&18.76&16.82&24.03&16.48\\ 
&20&20.91&21.63&23.33&20.96&27.40&18.70\\ 
&25&24.01&25.39&27.35&25.54&38.48&22.24\\ 
\hline
&5&11.74&11.80&11.95&12.12&11.89&11.80\\ 
&10&19.08&19.37&20.51&20.85&15.70&17.23\\ 
$\theta = 1.60$&15&25.50&26.61&28.57&25.91&35.98&23.16\\ 
&20&31.61&33.56&36.23&33.26&42.92&28.75\\ 
&25&37.92&40.27&43.32&40.71&57.52&34.58\\ 
\hline
&5&14.83&15.02&15.25&15.50&14.24&13.89\\ 
&10&25.57&26.38&28.03&28.40&22.36&23.21\\ 
$\theta = 1.80$&15&35.43&37.02&39.73&36.80&48.79&32.03\\ 
&20&42.66&44.79&48.32&47.47&58.47&38.99\\ 
&25&51.04&53.87&58.19&56.14&74.20&46.86\\ 
\hline
&5&17.31&17.48&17.81&18.11&16.92&16.33\\ 
&10&31.92&32.89&34.73&35.09&29.12&28.44\\ 
$\theta = 2.00$&15&44.33&46.19&49.66&46.76&60.48&39.85\\ 
&20&54.64&57.50&62.08&60.05&71.04&49.25\\ 
&25&63.46&66.89&71.67&71.04&85.97&58.42\\ 
\hline
\end{tabular}
\end{center}
}
\end{table}

\begin{table}[h]
\footnotesize{
\caption{Estimated power for the LFR alternative for small sample size}
\begin{center}
\begin{tabular}[c]{|c|c|c|c|c|c|c|c|}
\hline
&Sample&\multicolumn {3}{|c|}{Anis and Mitra (2011)} & Hollander and &  Fernandez-Ponce & Belzunce \\
LFR($\theta$)&size&\multicolumn {3}{|c|}{$T_0(j)$}&Proschan (1975)&et al. (1996)&et al. (2000)\\
&$n$&$j=0.25$&$j=0.50$&$j=1$&$T_1$&$T_5$&$T_6$\\
\hline
&5&6.85&6.94&7.04&7.13&6.02&6.54\\ 
&10&8.57&8.46&8.48&8.61&5.40&8.48\\ 
$\theta = 0.25$&15&10.12&10.01&9.80&8.79&11.22&9.92\\ 
&20&11.53&11.50&11.47&10.36&11.47&11.78\\ 
&25&13.17&12.75&12.67&12.98&15.42&13.54\\ 
\hline
&5&8.51&8.45&8.15&8.29&7.79&8.35\\ 
&10&11.57&11.59&11.70&11.87&7.25&11.62\\ 
$\theta = 0.50$&15&14.82&14.82&14.55&13.37&15.47&15.05\\ 
&20&18.06&17.91&17.58&16.81&16.97&17.98\\ 
&25&21.74&21.53&20.93&20.98&24.24&22.39\\ 
\hline
&5&8.99&9.00&9.12&9.24&8.80&9.24\\ 
&10&14.35&14.33&14.20&14.47&9.32&13.87\\ 
$\theta = 0.75$&15&19.03&19.04&19.03&17.60&19.53&18.73\\ 
&20&24.70&24.61&24.29&22.79&21.50&23.91\\ 
&25&29.15&28.97&28.40&28.02&28.87&28.65\\ 
\hline
&5&10.87&10.83&10.65&10.79&8.94&9.46\\ 
&10&16.75&16.72&16.54&16.85&11.26&16.37\\ 
$\theta = 1.00$&15&22.38&22.27&21.97&21.64&22.24&22.31\\ 
&20&28.94&28.87&28.48&27.14&25.77&27.94\\ 
&25&35.01&34.94&34.01&33.58&35.33&35.11\\ 
\hline
&5&11.70&11.61&11.68&11.91&10.03&10.88\\ 
&10&19.17&19.03&19.12&19.39&12.53&18.26\\ 
$\theta = 1.25$&15&26.09&26.07&25.86&23.91&26.57&26.46\\ 
&20&33.09&33.08&32.57&31.41&29.38&32.03\\ 
&25&41.19&41.09&40.22&37.65&40.11&40.67\\ 
\hline
\end{tabular}
\end{center}
}
\end{table}

\begin{table}[h]
\scriptsize{
\caption{Estimated power for the Weibull alternative for large sample size}
\begin{center}
\begin{tabular}[c]{|c|c|c|c|c|c|c|c|c|c|c|c|}
\hline
&&\multicolumn {3}{|c|}{Anis and Mitra} & Hollander \& & Koul & de Souza& Aly& Belzunce & Belzunce&Mudgadi\\
Weibull&Sample&\multicolumn {3}{|c|}{ (2011)}&Proschan&(1978)&Borges et&(1990)&et al. &et al.&\& Ahmed\\
($\theta$)&Size&\multicolumn {3}{|c|}{$T_0(j)$}&(1975)&&al. (1984)&&(2000)&(2001)&(2005)\\
&$n$&$j=0.25$&$j=0.50$&$j=1$&$T_1$&$T_2$&$T_3$&$T_4$&$T_6$&$T_7$&$T_8$\\
\hline
&30&13.46&13.49&13.81&13.01&6.79&8.04&14.98&12.33&12.40&1.11\\
&40&14.86&15.55&16.25&15.91&8.98&10.46&18.49&15.45&14.89&2.47\\
$\theta = 1.1$&50&17.25&18.15&18.81&17.64&10.61&12.53&19.51&16.78&16.73&3.96\\
&75&26.84&28.53&30.67&23.52&15.19&17.11&24.17&26.17&22.42&8.39\\
&100&31.53&33.32&35.27&29.33&19.65&21.78&28.84&30.65&27.67&13.06\\
\hline
&30&26.36&27.21&28.68&27.93&15.21&18.29&30.73&25.30&25.91&3.20\\
&40&33.57&35.37&36.85&35.34&20.55&24.80&37.17&31.44&32.54&8.14\\
$\theta = 1.2$&50&38.84&40.48&42.34&42.38&26.20&30.75&43.40&37.51&38.94&13.59\\
&75&59.90&62.65&65.96&57.82&40.39&45.50&56.17&57.78&54.19&28.61\\
&100&69.25&72.04&75.02&68.68&51.65&56.11&66.53&67.17&64.43&41.92\\
\hline
&30&44.94&46.38&48.59&46.93&27.13&32.82&48.92&41.76&43.58&8.45\\
&40&54.94&56.93&59.53&59.59&38.96&46.13&60.55&53.79&55.59&20.23\\
$\theta = 1.3$&50&65.10&67.81&70.42&68.82&48.28&55.93&68.59&62.31&64.26&31.68\\
&75&86.29&88.24&90.24&85.62&69.30&75.29&84.28&84.74&82.00&59.08\\
&100&92.84&94.15&95.44&93.17&81.99&86.39&91.82&91.72&90.64&77.28\\
\hline
&30&63.98&65.89&68.18&66.44&42.34&51.87&68.49&60.3&62.39&17.59\\
&40&75.49&77.42&79.83&79.31&57.81&66.60&79.87&73.21&75.05&37.23\\
$\theta = 1.4$&50&84.53&86.42&88.10&87.80&69.46&77.51&87.26&82.27&84.52&55.24\\
&75&97.31&97.98&98.40&97.19&88.89&92.88&96.82&96.58&95.62&83.62\\
&100&99.10&99.44&99.58&99.33&96.37&97.49&99.03&98.86&98.84&94.51\\
\hline
&30&78.30&80.18&81.87&81.58&58.39&68.59&83.49&75.35&77.39&30.73\\
&40&89.37&90.84&92.06&91.91&75.21&83.33&91.89&87.82&89.42&57.60\\
$\theta = 1.5$&50&94.57&95.63&96.53&96.24&84.96&90.79&96.35&93.40&94.34&75.69\\
&75&99.57&99.70&99.83&99.64&97.00&98.51&99.55&99.46&99.22&95.58\\
&100&99.93&99.95&99.98&99.96&99.63&99.77&99.93&99.91&99.89&99.42\\
\hline
\end{tabular}
\end{center}
}
\end{table}

\begin{table}[h]
\scriptsize{
\caption{Estimated power for the Gamma alternative for large sample size}
\begin{center}
\begin{tabular}[c]{|c|c|c|c|c|c|c|c|c|c|c|c|}
\hline
&&\multicolumn {3}{|c|}{Anis and Mitra} & Hollander \& & Koul & de Souza& Aly& Belzunce & Belzunce&Mudgadi\\
Gamma&Sample&\multicolumn {3}{|c|}{ (2011)}&Proschan&(1978)&Borges et&(1990)&et al. &et al.&\& Ahmed\\
($\theta$)&Size&\multicolumn {3}{|c|}{$T_0(j)$}&(1975)&&al. (1984)&&(2000)&(2001)&(2005)\\
&$n$&$j=0.25$&$j=0.50$&$j=1$&$T_1$&$T_2$&$T_3$&$T_4$&$T_6$&$T_7$&$T_8$\\
\hline
&30&14.50&14.89&15.67&14.39&7.46&8.64&6.38&12.86&13.32&1.32\\
&40&15.72&16.51&17.33&16.52&9.28&10.43&7.26&14.30&14.98&2.49\\
$\theta = 1.2$&50&18.14&19.13&20.72&20.04&11.67&12.97&8.77&17.22&18.62&4.38\\
&75&27.41&29.63&33.10&25.95&16.37&16.80&11.12&26.22&23.33&8.31\\
&100&32.16&34.26&38.13&31.52&20.99&22.04&14.75&30.37&28.39&13.41\\
\hline
&30&26.95&28.57&31.11&29.75&16.32&18.45&5.37&25.07&26.68&3.64\\
&40&33.07&35.22&38.48&38.31&22.67&25.02&6.95&31.11&34.09&8.33\\
$\theta = 1.4$&50&37.26&40.24&44.52&44.40&28.26&30.11&9.40&36.19&39.93&13.94\\
&75&56.62&60.78&66.78&58.25&42.01&41.98&15.91&54.20&52.98&26.59\\
&100&68.56&72.91&78.40&72.59&55.17&53.96&23.36&65.79&66.63&39.61\\
\hline
&30&42.27&44.77&49.04&48.29&28.20&32.08&3.95&39.71&42.71&8.45\\
&40&52.05&55.48&60.83&59.98&39.33&42.62&6.06&49.50&54.25&18.30\\
$\theta = 1.6$&50&59.84&63.92&69.77&68.61&49.36&51.06&8.60&57.54&62.90&28.84\\
&75&82.09&86.10&90.19&86.15&70.99&68.87&18.92&79.69&81.16&52.36\\
&100&89.64&92.81&95.79&93.43&84.39&79.66&32.11&88.17&89.87&68.77\\
\hline
&30&57.28&60.53&65.60&64.49&40.72&46.42&2.14&54.21&58.70&15.72\\
&40&68.78&72.96&78.04&78.45&57.74&60.30&4.17&67.11&72.24&32.58\\
$\theta = 1.8$&50&78.61&82.66&87.13&86.35&70.81&69.79&6.68&75.24&81.36&48.11\\
&75&93.59&95.98&97.88&96.32&89.35&85.14&19.70&92.13&93.66&74.16\\
&100&97.72&98.76&99.54&99.11&96.49&93.17&35.02&97.11&98.17&87.50\\
\hline
&30&70.68&74.26&79.04&78.67&55.26&59.35&1.07&66.47&71.81&24.39\\
&40&81.85&85.50&89.63&89.51&72.96&74.35&2.31&79.48&84.63&47.24\\
$\theta = 2.0$&50&89.37&92.27&95.07&94.91&84.29&83.04&4.99&87.00&91.81&64.06\\
&75&98.16&99.00&99.70&99.19&96.65&94.27&17.31&97.76&98.25&87.98\\
&100&99.70&99.87&99.96&99.95&99.37&98.27&36.28&99.64&99.80&96.47\\
\hline
\end{tabular}
\end{center}
}
\end{table}

\begin{table}[h]
\scriptsize{
\caption{Estimated power for the LFR alternative for large sample size}
\begin{center}
\begin{tabular}[c]{|c|c|c|c|c|c|c|c|c|c|c|c|}
\hline
&&\multicolumn {3}{|c|}{Anis and Mitra} & Hollander \& & Koul & de Souza& Aly& Belzunce & Belzunce&Mudgadi\\
LFR&Sample&\multicolumn {3}{|c|}{ (2011)}&Proschan&(1978)&Borges et&(1990)&et al. &et al.&\& Ahmed\\
($\theta$)&Size&\multicolumn {3}{|c|}{$T_0(j)$}&(1975)&&al. (1984)&&(2000)&(2001)&(2005)\\
&$n$&$j=0.25$&$j=0.50$&$j=1$&$T_1$&$T_2$&$T_3$&$T_4$&$T_6$&$T_7$&$T_8$\\
\hline
&30&14.52&14.68&14.49&14.03&7.12&9.20&27.05&15.23&13.74 &1.42\\
&40&18.28&17.99&17.75&17.37&9.77&12.45&30.96&18.44&17.03 &3.01\\
$\theta = 0.25$&50&22.02&21.68&20.82&20.45&12.60&16.09&35.55&21.59&19.96 &5.40\\
&75&35.99&36.11&35.06&27.30&18.27&23.86&44.89&35.07&26.97 &12.33\\
&100&42.70&42.43&40.70&33.39&23.84&31.78&52.82&41.59&32.69 &19.67\\
\hline
&30&25.14&24.90&24.37&23.30&12.59&16.42&48.94&24.73&22.96 &2.65\\
&40&30.84&30.79&30.29&31.08&18.47&24.54&58.11&32.86&30.57 &7.40\\
$\theta = 0.50$&50&39.09&38.88&37.14&36.13&23.00&30.38&64.26&38.31&35.52 &12.60\\
&75&60.54&60.54&58.65&50.37&35.67&47.29&77.42&59.35&49.07 &29.25\\
&100&72.38&71.60&69.13&62.00&46.72&60.99&86.33&70.47&60.60 &45.36\\
\hline
&30&34.24&33.85&32.77&31.22&17.78&23.48&66.20&33.36&31.00 &4.70\\
&40&44.04&43.72&42.32&40.55&25.43&33.35&74.40&42.92&39.97 &11.49\\
$\theta = 0.75$&50&51.21&51.03&48.74&49.47&32.30&43.60&81.43&51.68&48.33 &19.77\\
&75&75.50&75.25&73.53&65.55&49.24&63.37&91.33&74.40&64.14 &44.19\\
&100&85.72&85.41&83.39&77.97&63.36&77.48&95.92&84.82&76.54 &64.02\\
\hline
&30&41.50&41.32&40.23&39.06&22.12&30.14&77.66&40.54&38.52 &6.75\\
&40&52.14&51.66&50.06&49.68&32.02&42.43&85.22&51.55&48.53 &16.77\\
$\theta = 1.00$&50&62.41&62.16&59.58&58.91&41.11&53.73&90.12&61.58&57.59 &28.36\\
&75&84.92&84.67&82.81&76.59&60.63&74.31&96.35&83.86&75.48 &56.77\\
&100&92.50&91.97&90.46&86.88&75.08&86.98&98.74&91.64&85.54 &76.75\\
\hline
&30&48.14&48.30&47.22&44.58&27.06&35.31&84.82&46.06&43.89 &9.01\\
&40&59.20&59.37&57.48&56.67&37.33&49.08&91.19&58.65&55.36 &20.63\\
$\theta = 1.25$&50&69.38&69.08&67.02&66.81&48.79&61.33&94.23&69.08&66.00 &35.62\\
&75&90.07&89.85&88.40&83.26&68.38&81.94&98.65&89.32&82.32 &66.17\\
&100&95.72&95.41&94.22&92.13&82.82&92.24&99.50&95.09&91.32 &84.84\\
\hline
\end{tabular}
\end{center}
}
\end{table}

\end{document}